\documentclass[showpacs,preprintnumbers,amsmath,amssymb]{revtex4}

\usepackage{graphicx}
\usepackage{dcolumn}
\usepackage{bm}
\begin{document}
\preprint{Nuclear Theory}
\title{Single Pion production from Nuclei}
\author{S.K.Singh, M. Sajjad Athar and S. Ahmed  }
\email{pht13sks@rediffmail.com}
\affiliation{Department of Physics, Aligarh Muslim University, Aligarh-202 002, India.}
\date{\today}
 
\begin{abstract}
 We have studied charged current one pion production induced by $\nu_\mu(\bar\nu_\mu)$ from some nuclei. The calculations have been done for the incoherent pion production processes from these nuclear targets in the $\Delta$ dominance model and take into account the effect of Pauli blocking, Fermi motion and renormalization of $\Delta$ properties in the nuclear medium. The effect of final state interactions of pions has also been taken into account. The numerical results have been compared with the recent results from the MiniBooNE experiment for the charged current 1$\pi$ production, and also with some of the older experiments in Freon and Freon-Propane from CERN.
\end{abstract}
\maketitle 

The $\nu_\mu(\bar\nu_\mu)$ induced charged current production of pions from nuclear targets is calculated in a delta dominance model. In this model, the $\Delta$ resonance excited by $\nu_\mu(\bar{\nu_\mu})$ in the nuclear medium decays into pions, which undergo final state interaction with the nucleons in the nuclear medium. The main ingredients of the calculations are: model for the $\nu_\mu(\bar\nu_\mu)$ excitation of $\Delta$, $\Delta$ propagation and its decay and the final state interaction of pions in the nuclear medium. 

The weak excitation of nucleon to $\Delta$ is described in terms of vector and axial-vector transition form factors. The vector form factors are determined from the analysis of electromagnetic excitation of $\Delta$ for which recent data from JLab are used to fix the parameters of the form factors\cite{Lalakulich}. On the other hand, the axial vector form factor are determined from the older experiments done in hydrogen and deuterium targets. Once $\Delta$ is excited in the nucleus, its propagation is treated using relativistic Rarita Schwinger equation in terms of its mass and width which may change due to nuclear medium effects. These effects are included by calculating the self energy of $\Delta$ in nuclear medium using nuclear many body theory in a local density approximation. The real and imaginary part of the $\Delta$ self energy modify the mass and width of delta in the nuclear medium which are parameterized as a function of nuclear density\cite{Oset}. These $\Delta$s once produced in the medium decay to produce pions which undergo final state interaction with the nucleons in the medium and lose energy in the elastic scattering or get absorbed through inelastic processes. These are treated using Monte Carlo simulations by generating a pion of given momentum and charge at a point ${\bf r}$ in the nucleus. Assuming the real part of the pion nuclear potential to be weak compared with their kinetic energies, they are propagated following straight lines till they are out of the nucleus. At the beginning, the pions are placed at a point $({\bf r}={\bf b}, z_{in})$, where $z_{in}=-\sqrt{R^2-|{\bf b}|^2}$, with ${\bf b}$ as the random impact parameter, obeying $|{\bf b}|<$R. R is upper bound for the nuclear radius, which is chosen to be such that $\rho(R)\approx$10$^{-3}\rho_0$, with $\rho_0$ is the normal nuclear matter density. Then pion is made to move along the z-direction in small steps until it comes out of the nucleus\cite{Vicente}. 

In the following, we describe the formalism and present numerical results and discuss them. We also compare these results with the available experimental results.   

The cross section for neutrino induced charged current production of $\Delta^{++}$ on a free proton and its subsequent decay i.e.
\[\nu_\mu(k) + p \rightarrow \mu^-(k^{\prime}) + \Delta^{++}(p^{\prime}) \rightarrow \mu^- + p + \pi^{+}\]
is given by\cite{prd1}:
\begin{eqnarray}
\sigma(E_{\nu})=\frac{1}{128 \pi^2} \frac{M}{M_\Delta} \frac{G_F^{2}cos^2{\theta_c}}{(s-M^2)^2} \int_{q^2_{min}}^{q^2_{max}} dq^2
 \int_{k^{0}_{min}}^{k^0_{max}} {dk^0}^\prime~~ L_{\mu\nu} J^{\mu\nu}
 \frac{\Gamma(\bold{W})}{(\bold{W}-M_{\Delta})^2 +(\frac{\Gamma_{\Delta}}{2})^2}~~
\end{eqnarray}
where $s=(p+k)^2$, W is the $\Delta$ invariant mass, $ M(M_\Delta)$ is the nucleon(delta) rest mass, $\Gamma_\Delta$ is the $\Delta$ width, $L_{\mu\nu}$ is leptonic tensor and hadronic tensor $J^{\mu\nu}$ is defined in terms of the hadronic matrix element $J^\mu=\bar\Psi_\alpha A^{\alpha\mu} \Psi$, $\Psi_\alpha$ is Rarita Schwinger wave function for $\Delta$, $\Psi$ is the nucleon wave function, $A^{\alpha\mu}$ is the N-$\Delta$ transition vertex given  in terms of vector and axial vector $N-\Delta$ transition form factors which are taken from the work of Lalakulich et al.\cite{Lalakulich}.

When the above process takes place in the nuclear medium the width and mass of $\Delta$ are modified which is described in terms of the self energy $\Sigma$ as~\cite{Oset}:
\begin{eqnarray}
M_{\Delta} \rightarrow {\tilde M_{\Delta}}&=&M_{\Delta}+\mbox{Re}\Sigma_{\Delta}=
M_{\Delta} + 40 \frac{\rho}{\rho_{0}}MeV ~~and \nonumber\\
\frac{\Gamma}{2} \rightarrow  \frac{\tilde\Gamma}{2} - Im{{\Sigma}_{\Delta}}&=&\frac{\tilde\Gamma}{2} + [C_{Q}\left (\frac{\rho}{{\rho}_{0}}\right )^{\alpha}+C_{A2}\left (\frac{\rho}{{\rho}_{0}}\right )^{\beta}+C_{A3}\left (\frac{\rho}{{\rho}_{0}}\right )^{\gamma}]~~~~
\end{eqnarray}
where ${\tilde\Gamma}$ is Pauli corrected width. $C_{Q}$ accounts for the $\Delta N  \rightarrow
\pi N N$ process, $C_{A2}$ for the two-body absorption process $\Delta
N \rightarrow N N$ and $C_{A3}$ for the three-body absorption process $\Delta N N\rightarrow N N N$. The coefficients $C_{Q}$, $C_{A2}$, $C_{A3}$ and $\alpha$, $\beta$ and $\gamma$ are taken from Ref.~\cite{Oset}. We have taken energy dependent decay width for the $\Delta$.

The total scattering cross section for the neutrino induced charged current pion production process in the nucleus in the local density approximation is given by\cite{prd1}
\begin{eqnarray}
\sigma=\frac{G_F^{2}cos^2{\theta_c}}{256\pi^3}\int \int {d{\bf r}}\frac{d\bf{k^\prime}}{E_k E_{k^\prime}}\frac{1}{MM_\Delta}
\frac{\frac{\tilde\Gamma}{2}+C_{Q}\left (\frac{\rho}{{\rho}_{0}}\right )^{\alpha}}{(W- M_\Delta-Re\Sigma_\Delta)^2+(\frac{\tilde\Gamma}{2.}-Im\Sigma_\Delta)^2}
\left[\rho_p({\bf r})+\frac{1}{9}\rho_n({\bf r})\right]L_{\mu\nu}J^{\mu\nu}
\end{eqnarray}

\begin{figure}
  {\includegraphics[height=.3\textheight]{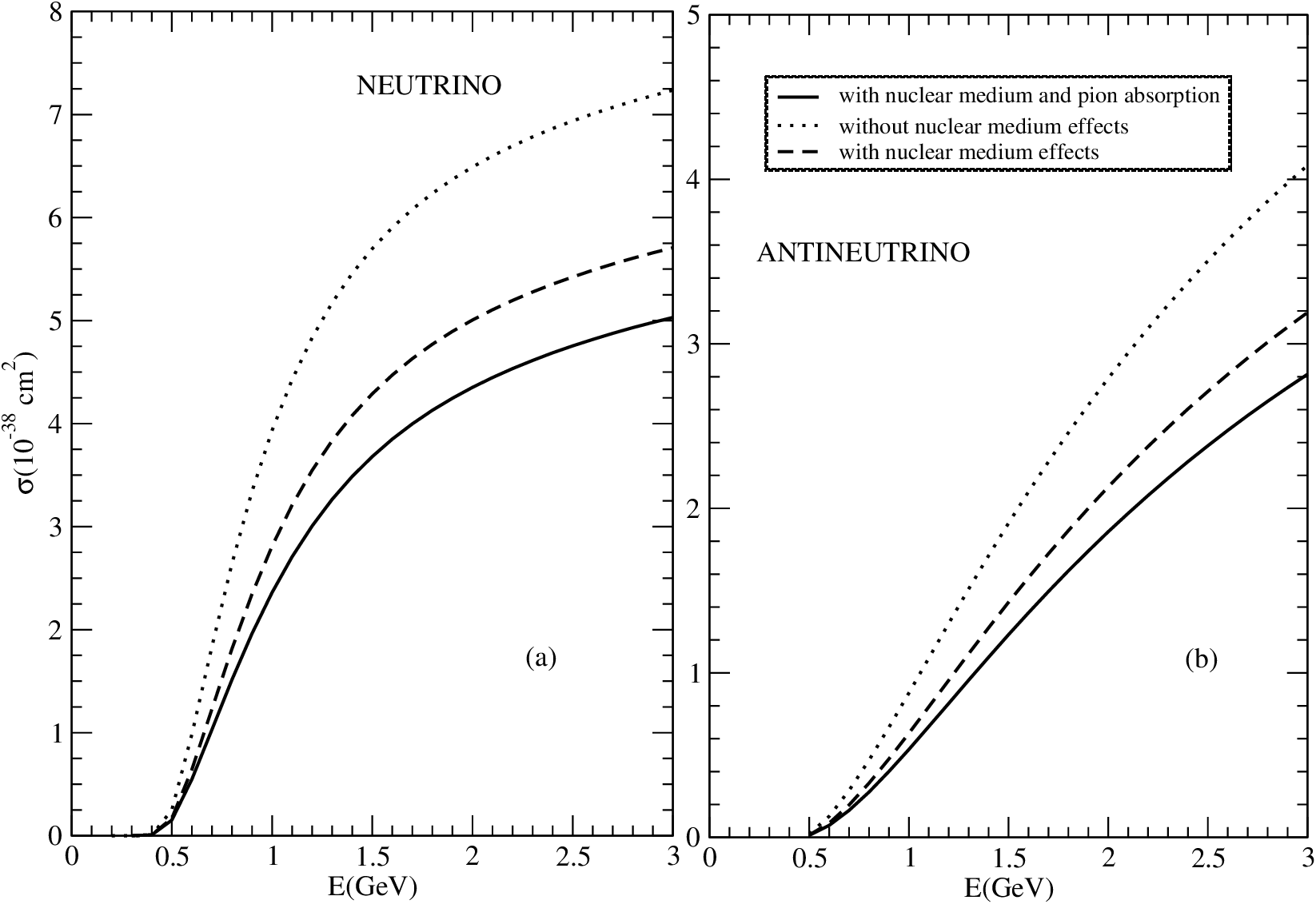}}
  \caption{(a) Charged current 1$\pi^+$ production cross section $\sigma$ vs $E_{\nu_\mu}$ for $^{12}C$. (b) Charged current 1$\pi^-$ production cross section $\sigma$ vs $E_{\bar\nu_\mu}$ for $^{12}C$.}
\end{figure}

\begin{figure}
  {\includegraphics[height=.3\textheight]{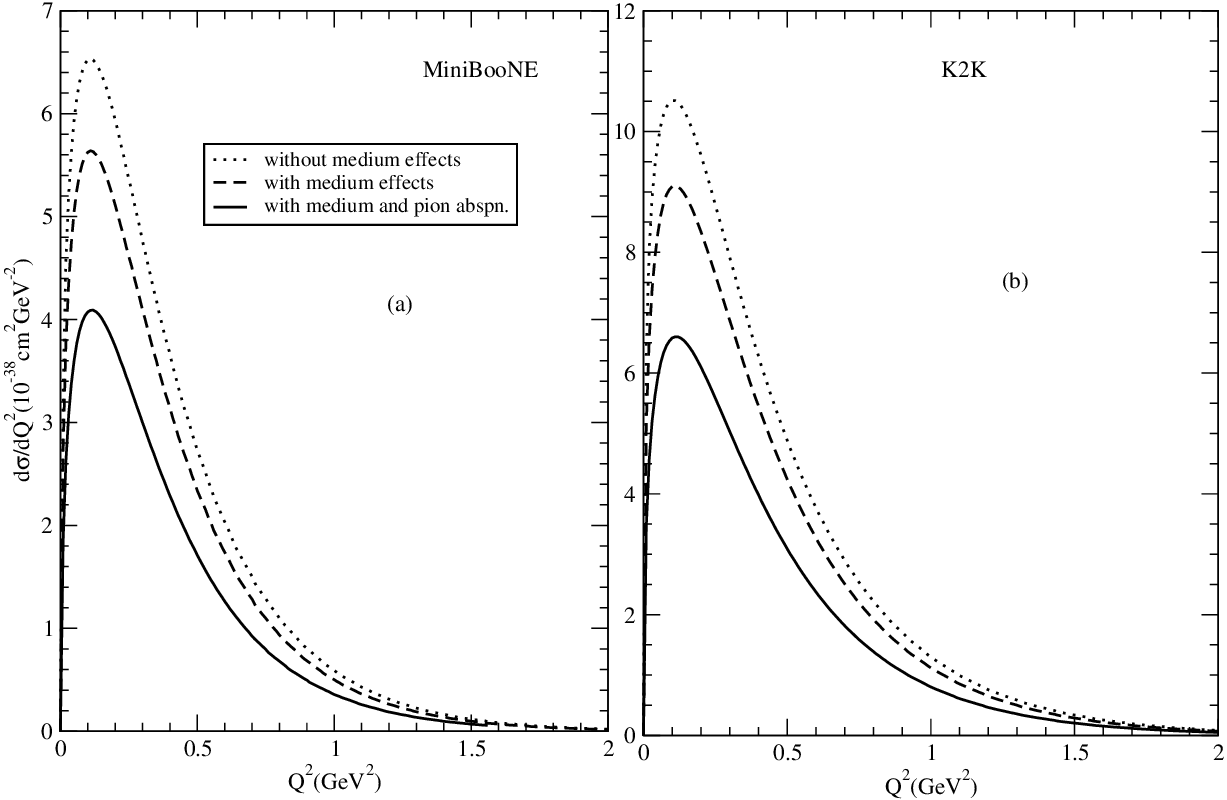}}
  \caption{$Q^2$ distribution averaged over MiniBooNE flux(Left panel) and K2K flux(right panel) for $\nu_\mu-^{12}C$ reaction.}
\end{figure}

\begin{figure}
  {\includegraphics[height=.3\textheight]{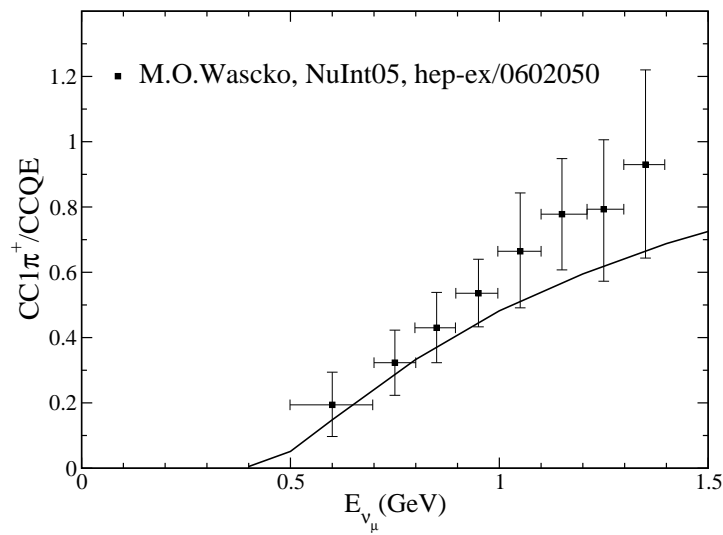}}
  \caption{$\frac{\sigma^{cc1\pi^+}}{\sigma^{CCQE}}$ in $^{12}C$.}
\end{figure}

\begin{figure}
  {\includegraphics[height=.3\textheight]{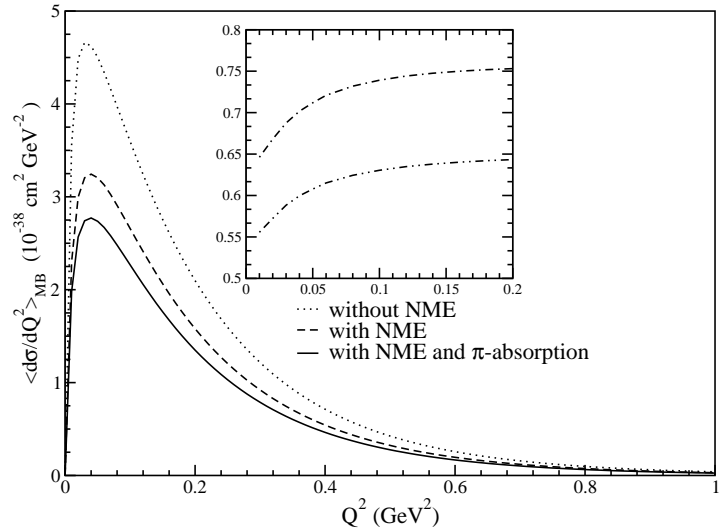}}
  \caption{$Q^2$ distribution averaged over MiniBooNE flux for $\bar\nu_\mu-^{12}C$ reaction. In the inset the ratios of the differential cross sections calculated including nuclear medium with(without) pion absorption effects to the cross section calculated without including nuclear medium effects have been shown.}
\end{figure}

\begin{figure}
  {\includegraphics[height=.3\textheight]{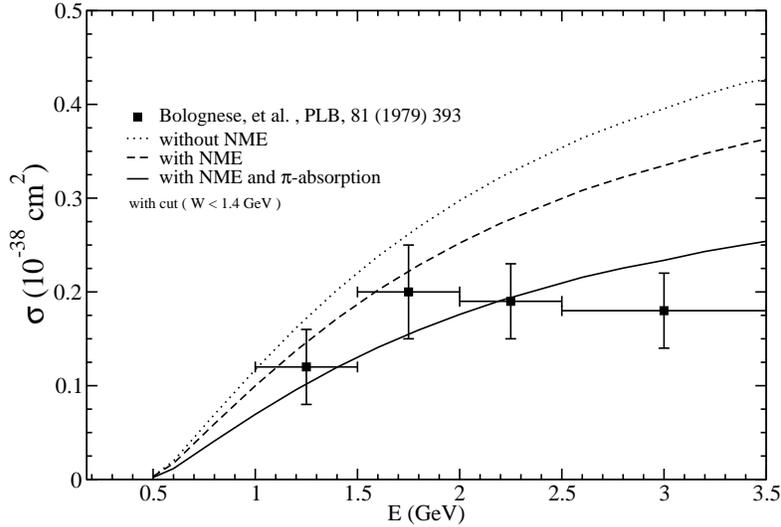}}
  \caption{Charged current 1$\pi^-$ production cross section $\sigma$ vs $E_{\bar\nu_\mu}$ for Freon-Propane.}
\end{figure}

The pions produced in this process are scattered and absorbed in the nuclear medium. This is treated in a Monte Carlo simulation using the  results of Vicente Vacas\cite{Vicente} for the final state interaction of pions.

In Fig.1, we show the total cross section for charged current single $\pi^+$(Fig.1a) and $\pi^-$(Fig.1b)
production from $^{12}C$ using the
N-$\Delta$ transition form factors given by Lalakulich et
al.\cite{Lalakulich}. We have presented the results for total scattering
cross section $\sigma(E_\nu)$ without the nuclear medium effects, with
the nuclear medium modification effects, and with nuclear medium and
pion absorption effects. For the incoherent process, we find that the
nuclear medium effects lead to a reduction of around 12-15$\%$ for
neutrino energies $\text E_\nu$=0.7-2GeV and when the pion absorption
effects are also taken into account along with the nuclear medium
effects the total reduction in the cross section is around
$30-40\%$.

In Fig.2, we have presented the results for the differential
scattering cross section $<\frac{d\sigma}{dQ^2}>$ vs $Q^2$ for
CC1$\pi^+$ production for the incoherent process
averaged over the MiniBooNE and K2K spectrum for $\nu_\mu$ induced
reaction in $^{12}C$ and $^{16}O$.

We have presented the ratio of the cross sections for charged current
1$\pi^+$(CC 1$\pi^+$) production to charged current quasielastic
scattering(CCQE) in Fig.3. For this purpose the cross section for
quasi-elastic charged lepton production is calculated in our
model\cite{singh} using weak nucleon axial vector and vector form
factors given by Bradford et al.(BBBA06)\cite{budd}. The results have been compared with the preliminary results from
MiniBooNE collaboration\cite{boone1}.

In Fig.4, we have presented the results for $\langle\frac{d\sigma}{dQ^2}\rangle$ vs $Q^2$ averaged over the MiniBooNE spectrum for anti-neutrinos. We find that for the $Q^2$ distribution, the reduction in the differential cross section with nuclear medium effects is around 15\% in the peak region of $Q^2$ which becomes 35\% when pion absorption effects are also included. 

In Fig.5, we present our results for the total cross section obtained for the incoherent 1$\pi^-$ production process on nucleon target induced by ${\bar\nu}_\mu$ on Freon-Propane($CF_3Br-C_3H_8$) and compare our results with the experimental results of Bolognese et al.~~\cite{bolognese}.

We have also studied the effect of various other~\cite{paschos1} parameterizations of N-$\Delta$ transition form factors on the differential cross section $\langle\frac{d\sigma}{dQ^2}\rangle$. We find that in the peak region of $\langle\frac{d\sigma}{dQ^2}\rangle$ the effect is quite small for antineutrino reaction but it could be 5-10\%  in the case of neutrino reactions\cite{prd1}.

To summarize, we have used the $\Delta$ dominance model to study nuclear medium effects in pion production processes induced by neutrinos and antineutrinos from nuclei at intermediate energies relevant to MiniBooNE and K2K experiments. We find that the nuclear medium effects like the modification of mass and width of delta in the nuclear medium and final state interaction of pions give an overall reduction of 15\% without pion absorption and 30\% with pion absorpton in the magnitude of total cross section and $Q^2$ distribution. However, the shape of $Q^2$ distribution is not affected by inclusion of these effects except at very low $Q^2$. The results for neutrino and antineutrino induced reactions are qualitatively similar.


\begin{thebibliography}{widest-label}
\bibitem{Lalakulich} O. Lalakulich, E. A. Paschos and G. Piranishvili, Phys. Rev. {\bf D 74}, 014009 (2006).
\bibitem{Oset} E. Oset and L. L. Salcedo, Nucl. Phys. {\bf A 468}, 631
 (1987); C. Garcia Recio, E. Oset, L. L. Salcedo, D. Strottman and
 M. J. Lopez, Nucl. Phys. {\bf A 526}, 685 (1991).
\bibitem{Vicente} M. J. Vicente Vacas, Private Communication.
\bibitem{prd1} M. Sajjad Athar, S. Ahmad and S. K. Singh, Phys. Rev. {\bf D 75}, 093003 (2007); Phys. Rev. {\bf D 74}, 073008 (2006).
\bibitem{singh} S. K. Singh and E. Oset, Phys. Rev. {\bf C 48}, 1246 (1993); ibid Nucl. Phys. {\bf A 542}, 587 (1992); M. Sajjad Athar, S. Ahmad and S.K.Singh, Eur. Phys. J. {\bf A 24}, 459 (2005).
\bibitem{budd} R. Bradford, A. Bodek, H. Budd and J. Arrington, Nucl. Phys. (Proc.Suppl.) {\bf B 159}, 127 (2006).    
\bibitem{boone1} M. O. Wascko, Nucl. Phys. {\bf B (Proc. Suppl.) 159}, 79 (2006). 
\bibitem{bolognese} T. Bolognese, J.P. Engel, J.L.Guyonnet and J.L.Riester, Phys. Lett. {\bf B 81}, 393 (1979).
\bibitem{paschos1}E. A. Paschos, J. Y. Yu and M. Sakuda,
 Phys. Rev. {\bf D 69}, 014013 (2004); P. A. Schreiner and F. von Hippel, Nucl. Phys. {\bf B 58}, 333 (1973).     
\end{thebibliography}
\end{document}